\documentclass[aps,prd,showpacs,superscriptaddress,twocolumn]{revtex4}

\usepackage{graphicx}
\usepackage{dcolumn}
\usepackage{bm}

\begin{document}

\def\be{\begin{equation}}
\def\ee{\end{equation}}
\def\bd{\begin{displaymath}}
\def\ed{\end{displaymath}}
\def\ba{\begin{eqnarray}}
\def\ea{\end{eqnarray}}
\def\lr{\leftrightarrow}
\def\B{\rm B}
\def\C{\rm C}
\def\D{\rm D}
\def\J{\rm J}
\def\K{\rm K}
\def\L{\rm L}
\def\P{\rm P}
\def\S{\rm S}
\def\x{\bf x}
\def\ss{$s\bar s$}
\def\cs{$c\bar s$}
\def\cc{$c\bar c$}

\title{\Large Options for the SELEX state D$_{s\J}^+(2632)$} 

\author{T.Barnes\footnote{Email: tbarnes@utk.edu}}
\affiliation{
Department of Physics and Astronomy, University of Tennessee,
Knoxville, TN 37996,
USA,\\
Physics Division, Oak Ridge National Laboratory,
Oak Ridge, TN 37831, USA}

\author{F.E. Close\footnote{Email: f.close@physics.ox.ac.uk}}
\affiliation{Rudolf Peierls Centre for Theoretical Physics, 
University of Oxford, Keble Rd., 
Oxford, OX1 3NP, United Kingdom}

\author{J.J.Dudek\footnote{Email: j.dudek1@physics.ox.ac.uk}}
\affiliation{Rudolf Peierls Centre for Theoretical Physics, 
University of Oxford, Keble Rd., 
Oxford, OX1 3NP, United Kingdom}

\author{S.Godfrey\footnote{Email: godfrey@physics.carleton.ca}}
\affiliation{Ottawa-Carleton Institute for Physics, 
Department of Physics, Carleton University, 
Ottawa K1S 5B6, Canada}

\author{E.S.Swanson\footnote{Email: swansone@pitt.edu}}
\affiliation{Department of Physics and Astronomy, 
University of Pittsburgh, 
Pittsburgh, PA 15260, USA}

\date{\today}

\begin{abstract}
We consider possible assignments for the D$_{s\J}^+(2632)$, which was
recently reported in D$_s^+\eta$ and D$^0$K$^+$ final states by the SELEX
Collaboration at Fermilab.  
The most plausible quark model assignment for this state 
is the first radial excitation ($2^3\S_1$) 
of the $c\bar s$ D$_s^*(2112)$, although
the predicted mass and strong decay branching fractions for 
this assignment are not in agreement with the SELEX data. 
The reported dominance of D$_s\eta$ over DK appears especially problematic.
An intriguing similarity to the K$^*(1414)$ is noted.
$2^3\S_1$--$^3\D_1$ configuration mixing is also considered, 
and we find that this effect is unlikely to resolve the branching fraction
discrepancy.
Other interpretations as a $c\bar s$-hybrid or a two-meson molecule
are also considered, but appear unlikely.
Thus, if this state is confirmed, it will require reconsideration of
the systematics of charmed meson spectroscopy and strong decays.
\end{abstract}


\maketitle

\section{Introduction}

The SELEX Collaboration 
\cite{Evdokimov:2004iy}
recently reported evidence for a new
charm-strange meson, known as the D$_{s\J}^+(2632)$, in
the final states D$_s^+\eta$ and D$^0$K$^+$. The strongest evidence
for the state is in D$_s^+\eta$, from which SELEX quote a mass of
M$ = 2635.9 \pm 2.9$~MeV. The D$^0$K$^+$ channel shows weaker evidence
for a similar state, with a mass and total width upper limit of
M$ = 2631.5 \pm 1.9$~MeV, $\Gamma < 17 \ {\rm MeV}$, $90\% \ {\it c.l.}$ 
These results led SELEX to a combined mass and total width limit of 
\be
{\rm M} = 2632.6 \pm 1.6 \ {\rm MeV}
\ee
and
\be 
\Gamma < 17 \ {\rm MeV}, \ \ 90\% \ {\it c.l.} 
\ee
The estimated branching fraction ratio for the two observed modes is
\be
{\rm B.F.}
(\D_{s\J}^+(2632) \to D^0K^+/ D_s^+\eta)
= 0.16 \pm 0.06 \ . 
\ee
Assuming that the observation of a state decaying strongly into 
at least one of these modes is correct, this implies the existence of 
a relatively narrow resonance with a minimum quark content of $c\bar s$ 
and natural spin-parity. 

If we restrict our initial consideration to 
conventional $c\bar s$ quark model states which were predicted by 
Godfrey and Isgur \cite{Godfrey:xj} to lie within 200~MeV of the
reported D$_{s\J}^+(2632)$ mass, we find only one possible 
assignment which does not currently have an experimental candidate, 
the 2$^3$S$_1$ radial excitation of the $c\bar s$ D$^*_s(2112)$. 
(The 2$^1$S$_0$ state is excluded by parity. The $^3\P_2$ state
is consistent with the signal seen by SELEX
at 2570~MeV in $D^0K^+$, which is
absent in their $D_s\eta$ data;
this is in accord with expectations 
from phase space for such a state.)
Our rather generous mass constraint is
motivated by the recent observation of the  
D$_{s\J}^+(2317)$  
\cite{Aubert:2003fg},
which lies about 150~MeV below quark model expectations.
The SELEX state is similarly about 100~MeV below the 
Godfrey-Isgur prediction 
of 2.73~GeV for the 2$^3$S$_1$ $c\bar s$ state. 
Other quark model predictions for the $2^3\S_1$ $c\bar s$ mass 
are in the range 2.71-2.76~GeV 
\cite{Gupta:1994mw,Zeng:1994vj,Ebert:1997nk,Lahde:1999ih}, with
one prediction of 2.81~GeV 
\cite{DiPierro:2001uu}.
The next closest natural spin-parity $c\bar s$ state 
with no experimental candidate is the $^3\D_1$, which is predicted 
at 2.90~GeV by Godfrey and Isgur, 
about 270~MeV above the D$_{s\J}^+(2632)$.
Thus $2^3\S_1$ $c\bar s$ appears to be the most 
plausible quarkonium assignment for the D$_{s\J}^+(2632)$. 
(This conclusion has been reached independently by Chao\cite{Chao:2004nb}.)
The decay pattern of this state could also in principle
be significantly modified by $2^3\S_1$--$^3\D_1$ configuration
mixing, which we will consider in our discussion.

Although the 
predicted Godfrey-Isgur mass of 2.73~GeV for the $2^3\S_1$ $c\bar s$ state 
is somewhat higher than observed for the D$_{s\J}^+(2632)$, we note
that several recent candidates for light $q\bar q$ radial 
excitations also have rather lower masses than predicted by this model.
Examples include the 2P $n\bar n$ ($n=u,d$) candidates 
$a_1(1640)$ and $a_2(1700)$ \cite{PDG2004},
which were predicted to be at 1820~MeV.
It may be that this model overestimates the energy gap for 
radial excitation of mesons with small reduced $q\bar q$ mass;
these states may be displaced in mass by additional non-valence effects
such as mixing with the two-meson continuum. 

In the following discussion we will consider the implications of
$2^3$S$_1$ $c\bar s$ quark model, hybrid, and molecular assignments for the
D$_{s\J}^+(2632)$ in more detail. 
We will find that the mass and peculiar strong branching fractions 
reported for this state appear inconsistent with any of these assignments. 

\section{The $\D_{s\J}^+(2632)$ and the $\K^*(1414)$}

It is notable that the $\D_{s\J}^+(2632)$ shares several common features
with the problematic excited strange vector $\K^*(1414)$
\cite{Barnes:2002mu}. The mass of the $\K^*(1414)$ appears too light
for a strange partner of the 2S $n\bar n$ candidates 
$\omega(1419)$ and $\rho(1465)$; these states are all roughly degenerate,
unlike the 1S $\K^*(892)$, $\omega(782)$ and $\rho(770)$.
This is illustrated
in Fig.1, which shows the striking similarity between the $n\bar q$ and 
$c\bar q$ spectra.
The strong decay modes of the $\K^*(1414)$ are also 
in disagreement with theoretical expectations.
The $^3\P_0$ decay model (see section \ref{dec}) predicts comparable 
branching fractions to 
$\pi \K$, $\pi \K^*$, $\eta \K$ and $\rho \K$ \cite{Barnes:2002mu}. 
The LASS Collaboration \cite{Aston:1984ab} however
reports a dominant $\pi \K^*$ 
mode, a weak $\pi \K$ mode (B.F. ca. 7\%), and only an 
upper limit for $\rho \K$, 
$\Gamma(\rho \K)/\Gamma(\pi \K) < 0.17$, $95\% \ c.l.$. The 
weak $\rho \K$ mode is especially surprising, since this branching fraction
should equal
$\pi \K^*$ modulo phase space differences.

\begin{figure}[t]
\includegraphics[width=8cm,angle=0]{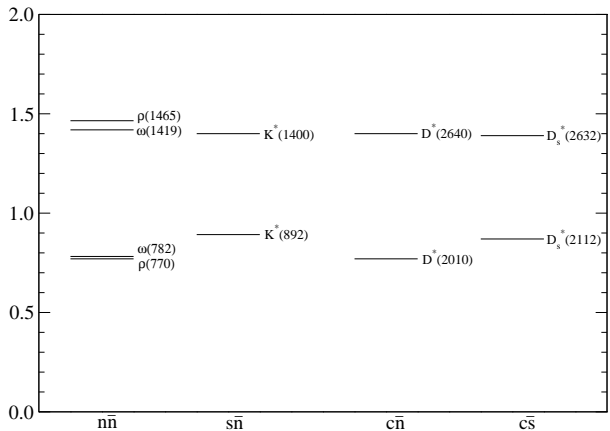}
\caption{
The experimental spectrum of 
1S and 2S vector quarkonium candidates,
contrasting states with $n$ and $c$ quarks  
combined with $\bar n$ and $\bar s$ antiquarks. 
The charmed spectrum is shifted downwards 
by 1.238~GeV for comparison.
Note that the 2S strange candidates
K$^*(1414)$ and D$_{s\J}^+(2632)$
are roughly degenerate with their nonstrange partners.}
\label{expt}
\end{figure}

Although these discrepancies do not explain the nature of the 
$\D_{s\J}^+(2632)$, they do suggest that the 
$\K^*(1414)$ and $\D_{s\J}^+(2632)$ may be closely related; 
both are lighter than expected given their nominal
nonstrange 2S partner radial excitations, and both show patterns 
of strong decays that differ considerably from 
the expectations of the $^3\P_0$ model. Thus, what is learned 
about one state may be useful in understanding both.

\section{Quark Model $c\bar s$ Interpretation\label{dec}}

As stated in the introduction, the only plausible $c\bar s$ assignment
for the D$_{s\J}^+(2632)$ is $n^{(2\S+1)}\L_{\J} = 2^3\S_1$, 
which has a predicted mass in the Godfrey-Isgur model of 2730~MeV. 
Allowed open-flavor decay modes for this state, assuming the SELEX mass 
of 2632~MeV, are DK, D$_s\eta$ and
D$^*$K.  The first two modes have two 1S pseudoscalars in the final state, and
hence are related by flavor matrix elements. This relation is
${\cal A}(\D_s^+\eta) = \sin\theta\, {\cal A}(\D^0\K^+)$, where ${\cal A}$ is a strong
decay amplitude and $\sin\theta \approx -1/\sqrt{2}$ is the amplitude of the $s\bar s$
component of the $\eta$. Assuming the $^3\P_0$ decay model and identical
D and D$^*$ spatial wavefunctions, the
decay amplitude to D$^*$K is also proportional to the same function, 
${\cal A}(\D^*\K) = -\sqrt{2}\; {\cal A}(\D\K)$. 
Thus, one expects {\it reduced} relative strong 
decay widths 
(summed over charge modes, but divided by the momentum-dependent decay
amplitude squared) of
$\D^*\K  : \ {\rm DK}\ :\ {\rm D}_s \eta \  = \ 4\ :\ 2\ :\ 1 \ $. 

As a simple initial estimate of physical branching fractions, 
since these are all P-wave decays we may assume a $p_f^3$ threshold dependence
for all modes, which gives expected relative branching fractions (again summed
over all charge modes) of
\be
{\rm B.F.}
\; (\D^*\K\ : \ {\rm DK}\ :\ {\rm D}_s \eta ) \  
= \ 4.2 \ :\ 7.0 \ :\ 1  \ . 
\ee
This is clearly in disagreement with the SELEX result
(assuming equal D$^0$K$^+$ and D$^+$K$^0$ modes) of
\be
{\rm B.F.}
\; ({\rm DK}\ :\ {\rm D}_s \eta ) \  = \ 0.32 \pm 0.12\ :\ 1  \ . 
\ee

These simple phase space arguments can sometimes be misleading, especially 
for radially excited states.
A familiar example is provided by the relative branching fractions of the 
$3^3$S$_1$ $c\bar c$ meson $\psi(4040)$ to
$\D\bar {\D}$, $\D \bar {\D}^* + h.c. $ and $\D^*\bar {\D}^*$.
Spin counting rules lead to expected relative branching fractions 
of 1:4:7 for these modes.   
This simple estimate however is invalidated by a node in the 
$^3$P$_0$-model $\D\bar {\D}$ decay amplitude near the physical point
\cite{LeYaouanc:1977ux}, which strongly suppresses the $\D\bar {\D}$ width, 
in agreement with experiment.

In view of the possible complication of nodes in the strong decay amplitudes of 
radially excited vector mesons, we have evaluated these amplitudes 
for the 
D$_{s\J}^+(2632)$
in the $^3$P$_0$ decay model 
\cite{LeYaouanc:1977ux,Micu:1968mk,LeYaouanc:1972ae,LeYaouanc:1973xz,
LeYaouanc:1974ri,LeYaouanc:1974mr,Ackleh:1996yt}, 
given a
$2^3$S$_1$
$c\bar s$ 
assignment.
The $^3$P$_0$ model, which assumes that strong decays proceed through local
$q\bar q$ pair creation with vacuum quantum numbers,
is the standard quark model approach for estimating 
strong decay widths. It has proven quite successful in describing a 
wide range of meson \cite{Barnes:1996ff,Barnes:2002mu}
and baryon \cite{Capstick:2000qj,Capstick:kb} decays.

\begin{figure}[t]
\includegraphics[width=6cm,angle=270]{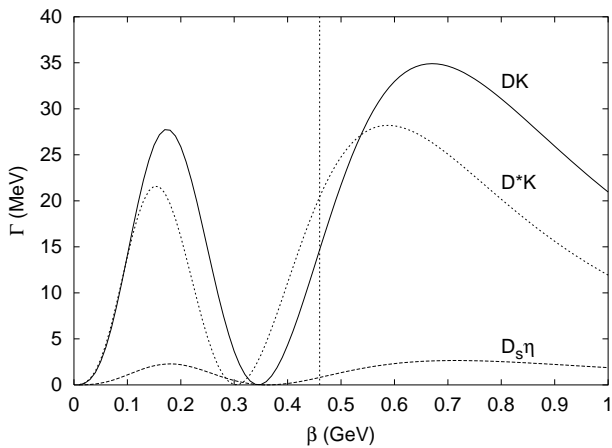}
\caption{
Theoretical partial widths of the $\D_{s\J}^+(2632)$ into 
$\D^*\K$, $\D\K$ and $\D_s \eta$ final states
in the $^3$P$_0$ decay model, assuming a
$2^3\S_1$ $c\bar s$ assignment. 
The widths are shown as functions of the $\D_{s\J}^+(2632)$ SHO
width parameter $\beta$ ($\beta \approx 0.46$~GeV is preferred theoretically).
See text for other parameter values.
}
\label{ws}
\end{figure}

The pair creation strength used here is $\gamma = 0.4$, which gives
reasonable numerical strong widths for a wide range of 
$n\bar n$, $n\bar s$, $s\bar s$ and 
$c\bar c$ mesons \cite{Barnes:1996ff,Barnes:2002mu}.
We use simple nonrelativistic SHO wavefunctions 
for all mesons. The wavefunction width parameter
$\beta$ is fixed separately for each meson to give the RMS radius 
predicted 
for that state
by the Godfrey-Isgur relativised quark model 
\cite{Godfrey:xj}.
This gives the values
$\beta(\eta_{s\bar s}) = 0.64$~GeV, $\beta(\K) = 0.61$~GeV, 
$\beta(\D) = 0.60$~GeV, 
$\beta(\D_s) = 0.65$~GeV 
and $\beta(\D^*) = 0.52$~GeV.
This procedure gives $\beta(\D_{s\J}^+(2632)) = 0.46$~GeV,
which is our preferred value. In the following 
we will treat this as a free parameter,
to explore the sensitivity of our results to wavefunction variation.
Masses of 330 MeV, 550 MeV, and 1600 MeV have been used for the
up, strange, and charm quarks respectively.

The resulting partial widths as functions of the 
$\D_{s\J}^+(2632)$ width parameter $\beta$
are shown in Fig.\ref{ws}. 
For width parameters near our preferred physical value
of $\beta = 0.46$~GeV
we find the branching fraction hierarchy
$\Gamma(\D^*\K) \agt $  
$\Gamma(\D\K) \gg $  
$\Gamma(\D_s\eta)$. This conclusion is complicated somewhat
by the presence of nodes near $\beta = 0.30 - 0.35$ GeV,
where there is a small total width and rapidly varying branching ratios. 
At our preferred value of $\beta$ 
the $\D_{s\J}^+(2632)$ total width 
is 36~MeV, and the ratio $\Gamma(\D\K)/\Gamma(\D_s \eta)$ is about 9. 
It is clear that the reported ratio
of $\Gamma(\D\K)/\Gamma(\D_s\eta)$ in Eq.(3) is not consistent with
the expectations of the $^3$P$_0$ decay model for 
$\D_{s\J}^+(2632)$ wavefunction width parameters near our
preferred $\beta$.

The prediction of comparable 
couplings of a $2^3\S_1$ $c\bar s$ quark model state at 2632~MeV
to DK and D$^*$K near our preferred $\beta$
is also evident in Fig.\ref{ws}.
A search for the D$^*$K mode and an accurate determination of the 
relative branching fractions of the $\D_{s\J}^+(2632)$ to 
D$^*$K, DK and $\D_s\eta$ should provide very useful tests of the 
$2^3\S_1$ $c\bar s$ assignment.

It is possible that 
$2^3\S_1$--$1^3\D_1$ quarkonium mixing may 
significantly alter the ratio of $\D\K$
and $\D_s\eta$ branching fractions.  
Such mixing may be generated by coupling to shared
virtual meson-meson states. We investigate this scenario by 
assuming a simple mixed state

\begin{equation}
|\D_{s\J}(2632)\rangle = \cos\theta |2^3\S_1\rangle + \sin\theta |1^3\D_1\rangle
\end{equation}
and examining the resulting ratio of $\D\K$ to $\D_s\eta$ widths. 
Direct computation in the
$^3\P_0$ model reveals that 
${\cal A}(^3\D_1\to \D\K)/{\cal A}(2^3\S_1\to \D\K)$ is very close
to ${\cal A}(^3\D_1\to \D_s\eta)/{\cal A}(2^3\S_1\to \D_s\eta)$. 
This may be expected since the same amplitude drives the 
$\D\K$ and $\D_s\eta$ modes.  Thus the ratio of $\D\K$ to
$\D_s\eta$ widths is very nearly independent of the mixing angle $\theta$. 
The exception is a narrow band around 

\begin{equation}
\theta = \tan^{-1}\left(-{{\cal A}(2^3\S_1\to \D\K)\over 
{\cal A}(^3\D_1 \to \D\K)}
\right) \approx 0.86\, \pi \ ,
\end{equation}
due to slightly offset nodes in the $\D\K$ and $\D_s\eta$ amplitudes.  
Suppression of the DK mode relative to 
$\D_s\eta$ therefore requires a mixing angle of 
$\theta \approx 155^{\circ}$ or $-25^{\circ}$.  
The near equality of the $^3\D_1$ to $2^3\S_1$ ratios mentioned above 
implies however that the absolute partial widths to 
$\D\K$ and $\D_s\eta$ are both very small near these specific mixing angles.
Thus one can only achieve a small $\D\K$ to $\D_s\eta$ branching fraction
ratio at the expense of considerably suppressing the absolute 
partial widths to these modes. It therefore
appears unlikely that $2^3\S_1$--$1^3\D_1$ $c\bar s$ mixing 
can explain the reported SELEX $\D\K / \D_s\eta $ branching fraction ratio.

One might also consider searches for radiative transitions from 
the $\D_{s\J}^+(2632)$. An estimate of the E1 radiative partial widths 
of a $2^3\S_1$ $c\bar s$ $\D_{s\J}^+(2632)$ to P-wave $\D_s$ states 
may be extracted from the results of Ref.\cite{Godfrey:2003kg}.
These rates are found to be quite small, typically $\alt 1$~keV,
so the $\D_{s\J}^+(2632)$ should not be visible in these radiative
modes with current statistics 
if it is indeed dominantly a $2^3\S_1$ $c\bar s$ state.

Similarly one can consider searches for closed-flavor dipion
hadronic transitions of the 
$\D_{s\J}^+(2632)$.  Following Ref.\cite{Godfrey:2004ya}, we estimate 
a partial width of 
$\Gamma(2^3\S_1 (c\bar{s})\to \D_s^* +\pi\pi) \simeq 220$~keV, 
which for $\Gamma_{tot.} < 17$~MeV implies a B.F. of $\agt 1$\%. 
It may therefore 
be possible to observe the $\D_{s\J}^+(2632)$ in this channel.  The 
analogous width of a $1^3\D_1$ state at this mass is estimated to be
$\Gamma(1^3D_1 (c\bar{s})\to \D_s^* +\pi\pi) \simeq 13$~keV which is 
probably too small to be observed. 

\section{Hybrid Assignment}

Hybrid mesons, in which the gluonic degree of freedom is excited,
should give rise to an ``overpopulation" of the hadron
spectrum relative to the expectations of the naive quark model.
In the meson spectrum hybrids may be identified by their exotic
J$^{\P\C}$ quantum numbers, provided that the mesons have definite
C-parity. The $\D_s$ sector however does not have definite C-parity,
so the spectrum of hybrids must be identified through the overpopulation
of states and the anomalous properties of these additional excitations.
The quantum numbers of the lightest $c\bar s$-hybrid multiplet in the
flux-tube model are J$^{\P}$ = $0^{\pm}, 1^{\pm}, 2^{\pm}$ which implies 
that overpopulation of the natural-J$^{\P}$ D$_s$ spectrum 
should first be evident in the $0^+$, $1^-$ and $2^+$ sectors.

As the $\D_{s\J}^+(2632)$ is reported to have strong decay branching 
fractions that differ from $^3$P$_0$ decay model expectations for 
the only likely $c\bar s$ candidate, it is natural to consider whether 
a $c\bar s$-hybrid assignment is plausible for this state. 
Unfortunately this interesting possibility does not appear to be consistent
with recent mass estimates for hybrids. Although the unequal $q$ 
and $\bar q$ mass case has not been considered in detail in the 
literature, $c\bar s$ is intermediate in quark mass between $c\bar c$ and
light $n\bar n$ hybrids, which have been studied using lattice gauge theory 
(LGT) and various
models. The flux-tube model \cite{Barnes:1995hc} finds a hybrid mass gap 
of M$_{\rm H}$ - M$_{\rm 1S} \approx $ 1.3~GeV for light 
$n\bar n$ quarks and 
$\approx $ 1.1~GeV for $c\bar c$. This is roughly consistent with
LGT studies 
\cite{Bernard:1997ib,Lacock:1998be,Luo:2002rz,Liao:2002rj},
which typically find hybrid mass gaps of 
M$_{\rm H}$ - M$_{\rm 1S} \approx $ 1.3~GeV 
for both $c\bar c$ and $n\bar n$ systems. 
Apparently the hybrid gap has 
little dependence on quark mass, which leads to an expected
$c\bar s$-hybrid mass of 3.2 -- 3.4~GeV. 

There are however 
experimental candidates for hybrids 
at rather lower masses, the best established of which is the 
$\pi_1(1600)$ \cite{Adams:1998ff}.
A recent quenched LGT study with light quarks \cite{Bernard:2003jd} 
also finds a somewhat smaller $n\bar n$ hybrid mass gap, consistent
with the $\pi_1(1600)$ being a hybrid. 
This suggests a hybrid mass gap of 1.0~GeV for
light quarks, and a $c\bar s$-hybrid mass of ca. 3.1~GeV.

In either case the expected hybrid mass is sufficiently far above the 
$\D_{s\J}^+(2632)$
mass to make this a very speculative possibility, which in our opinion
does not merit further consideration without evidence that the hybrid
mass gap is much lower than current theoretical expectations.

\section{Molecular Assignment}

The possibility that loosely bound states of mesons may 
exist in the charm sector
was first suggested many years ago \cite{Voloshin:1976ap,DeRujula:1976qd} 
in response to the reported anomalous strong decays of the 
$\psi(4040)$. Such ``molecular" meson bound states 
are allowed in principle in QCD; whether they actually do 
form in a given channel
is a question of detailed dynamics. Unfortunately, our current understanding of
interhadron forces is not sufficiently well developed to allow 
reliable predictions of the spectrum of hadronic molecules in general,
and the existing predictions tend to be rather model dependent.
Examples of hadron interaction models that anticipate molecular 
bound states in various channels are pion-exchange models 
\cite{Tornqvist:1993ng,Ericson:1993wy},
the constituent quark model 
\cite{Weinstein:1990gu,Dooley:1991bg}, 
and multiple gluon exchange models
\cite{Szczepaniak:2003vy}.

Since the residual interhadron forces that can lead to molecular
bound states are relatively weak, one would expect hadronic molecules 
to form most easily as S-wave bound states just below threshold.
Examples include the $f_0(980)$ and $a_0(980)$ just below 
$\K\bar {\K}$ threshold, which may be $\K\bar {\K}$ molecules 
\cite{Weinstein:1990gu};
the $\D_s(2317)$, which may be an analogous DK molecule 
\cite{Barnes:2003dj,vanBeveren:2004bz};
and the X$(3872)$, which may be a D$\bar {\rm D}^*\ + h.c.\ $
bound state 
\cite{Tornqvist:2003na,Close:2003sg,Braaten:2003he,Swanson:2003tb}.
In all cases these states have the quantum 
numbers of the two-meson pair in S-wave, and are at most 10s of
MeV below threshold.

A plausible meson molecule assignment for the 
$\D_{s\J}^+(2632)$
would similarly require a two-meson threshold at most
10s of MeV above the resonance mass, with S-wave quantum numbers
consistent with the $\D_{s\J}^+(2632)$. The only
two-meson system with the required quantum numbers of I = 0, 
natural J$^{\P}$, and quark content $c\bar s q\bar q$
that is within 100~MeV of the $\D_{s\J}^+(2632)$ 
is $\D_s^*\eta$, at a mass of 2660~MeV. Unfortunately
this system does not appear plausible for a molecular
bound state with $\D_{s\J}^+(2632)$ quantum numbers, 
since natural J$^{\P}$ would require the $\D_s^*\eta$
pair to be in a P-wave. This also
applies to all pseudoscalar-pseudoscalar and
pseudoscalar-vector pairs. A vector-vector pair would
give the lightest possible S-wave natural parity molecules,
but the lightest such systems with $\D_{s\J}^+(2632)$ 
quantum numbers are $\D_s^*\omega$ and $\D^*\K^*$, 
which are both close to 2.90~GeV. The required 
binding energy of 270~MeV appears implausibly large for a 
two-meson molecule.
  
We conclude that there are no two-meson systems with 
D$_{s\J}^+(2632)$ quantum numbers sufficiently nearby in mass
to admit an S-wave molecular bound state as a possible 
assignment for this resonance. 

\section{Multiquark Assignments}

More exotic possibilities can be considered
for the D$_{s\J}^+(2632)$, such as a 
$cq\bar s \bar q$ multiquark state
\cite{Maiani:2004xg,Chen:2004dy,Nicolescu:2004in,LK}.
Of course a multiquark system that is above a fall-apart
decay threshold would be expected to be extremely broad 
or nonresonant, and if the D$_{s\J}^+(2632)$ is a 
$cn\bar s \bar n$ or
$cs\bar s \bar s$ 
multiquark (for example) one would have to explain why the fall-apart
modes DK, D$^*$K and/or D$_s\eta$ do not make this an extremely broad state. 

One should note that there is a qualitative difference between molecule
and multiquark assignments, despite the fact that they share the same sector
of Hilbert space. Thus one might argue from quark content alone that the 
D$_{s\J}^+(2317)$ sets a scale of 2.32~GeV for the 
$c\bar{s}(u\bar{u} + d\bar{d})$ sector, and with an increase of 150~MeV
for each $s$ quark one could accommodate a $c\bar{s}s\bar{s}$ system
near the mass of the D$_{s\J}^+(2632)$. 
A $c\bar{s}s\bar{s}$ multiquark 
state might {\it a priori} have the large coupling
to D$_s\eta$ reported for the D$_{s\J}^+(2632)$. 
However, this is misleading because
the mass of the D$_{s\J}^+(2317)$ is actually determined by the DK threshold
if it is largely a DK molecular state, and there is no analogous S-wave
threshold that could explain the D$_{s\J}^+(2632)$. 
 
\section{Summary and Conclusions}

In this paper we have considered several possible assignments 
for the D$_{s\J}^+(2632)$ resonance recently reported by 
the SELEX Collaboration. Given the mass and allowed quantum numbers for
this state, the most plausible conventional $q\bar q$ quark model 
assignment is a $2^3$S$_1$ $c\bar s$ radial excitation of the 1S vector 
D$_s^*(2112)$. 
The mass reported by SELEX however is rather lower than predicted 
for this state, and the decay branching fractions disagree strongly 
with expectations. Theoretically, for a $2^3$S$_1$ $c\bar s$ state at
this mass one predicts a small D$_s\eta$ mode and comparable
DK and D$^*$K modes, with a total width of $\approx 36$~MeV. 
The SELEX Collaboration instead report a much larger branching fraction
to D$_s\eta$ than DK, contrary to expectations.
For this reason we find that it is difficult to accommodate the  
D$_{s\J}^+(2632)$ as a conventional $q\bar q$ mesons. 

Although the reported properties of the 
D$_{s\J}^+(2632)$ do not agree with quark model
expectations for a radially excited $2^3$S$_1$ $c\bar s$ vector, 
we noted that some of the unusual aspects of this state are
reminiscent of the 
strange vector meson K$^*(1414)$, which is also a $2^3$S$_1$
candidate. If the D$_{s\J}^+(2632)$ is confirmed,
a comparison of these two resonances may prove enlightening.

We also considered two other possible interpretations for the 
D$_{s\J}^+(2632)$, a $c\bar s$-hybrid and a two-meson charm-strange 
molecule. Both of these assignments appear very unlikely, given
the mass and quantum numbers reported for this resonance. 
We conclude that either our understanding of meson spectroscopy
and especially strong decays is inadequate to explain this state, 
or it is simply an experimental artifact. 

Future experimental studies will be crucial for understanding the
$\D_{s\J}^+(2632)$. 
The most important measurement (provided that the state is confirmed) 
will be the determination of the J$^{\P}$ quantum numbers,  
through the angular distributions of D$_{s\J}^+(2632)$ final 
states; this could support or eliminate our preferred $1^{-}$ 
(2$^3$S$_1$) assignment.  A large branching fraction to the $\D^*\K$ final 
state is another important prediction of the 2$^3$S$_1$ assignment,
which should be searched for. An accurate determination of the 
relative branching fractions to $\D^*\K$, $\D\K$ and $\D_s\eta$
is clearly of great importance, since this is where there is currently
evidence of disagreement with strong decay predictions for the 
2$^3$S$_1$ $c\bar s$ assignment.  
A 2$^3$S$_1$ $c\bar s$ state should also have a closed flavor dipion 
decay to $\D_s\pi\pi$ with a branching fraction of $\approx 1$\%, 
analogous to the decay $\psi' \to J/\psi \pi\pi$.
This may also be observable, especially in the high statistics environment of 
the B-factories. 
If the 2$^3$S$_1$ $c\bar s$ assignment is correct, there should 
be a second $1^{-}$ $c\bar{s}$ state ($^3$D$_1$) approximately 
200~MeV higher in mass; observation of this state would provide additional 
evidence in favor of a conventional $c\bar{s}$ interpretation that does not
rely on the predictions of strong decay models.
The observation of two new D$_s$ states in 
B meson decay and the fact that 
the $1^3$S$_1$ $\D_s^*$ state is produced in B~decay
with a branching fraction of several percent suggests that the 2S 
$\D_s^*$  
radial excitation should also be produced with a sizable branching fraction.  
We therefore expect that the $\D_{s\J}^+(2632)$ should also be
evident in B~decays, provided that it is indeed the
2$^3$S$_1$ $c\bar{s}$ state.

In view of the surprising properties reported for the D$_{s\J}^+(2632)$,
if confirmed it will require reconsideration of theoretical 
expectations for both the spectrum and the strong decay systematics of 
charmed mesons. Confirmation (or refutation) of the D$_{s\J}^+(2632)$ 
is clearly an important priority for meson spectroscopy.

\acknowledgments

This research was supported in part by the
U.S. National Science
Foundation through grant NSF-PHY-0244786 at the University of Tennessee,
the U.S. Department of Energy under contract DE-AC05-00OR22725 at
Oak Ridge National Laboratory (Barnes),
the Particle Physics and Astronomy Research Council, 
and the EU-TMR program ``Euridice'' HPRN-CT-2002-00311 (Close and Dudek)
the Natural Sciences and Engineering Research Council of Canada (Godfrey), 
and the U.S. Department of Energy under contract DE-FG02-00ER41135 (Swanson).


\begin{thebibliography}{99}

\bibitem{Evdokimov:2004iy}
A.~V.~Evdokimov  [SELEX Collaboration],
arXiv:hep-ex/0406045.

\bibitem{Godfrey:xj} 
S.~Godfrey and N.~Isgur, 
Phys.\ Rev.\ D {\bf 32}, 189 (1985).

\bibitem{Aubert:2003fg}
B.~Aubert {\it et al.}  [BABAR Collaboration],
Phys.\ Rev.\ Lett.\  {\bf 90}, 242001 (2003)
[arXiv:hep-ex/0304021].

\bibitem{Gupta:1994mw}
S.~N.~Gupta and J.~M.~Johnson,
Phys.\ Rev.\ D {\bf 51}, 168 (1995)
[arXiv:hep-ph/9409432].

\bibitem{Zeng:1994vj}
J.~Zeng, J.~W.~Van Orden and W.~Roberts,
Phys.\ Rev.\ D {\bf 52}, 5229 (1995)
[arXiv:hep-ph/9412269].

\bibitem{Ebert:1997nk}
D.~Ebert, V.~O.~Galkin and R.~N.~Faustov,
Phys.\ Rev.\ D {\bf 57}, 5663 (1998)
[Erratum-ibid.\ D {\bf 59}, 019902 (1999)]
[arXiv:hep-ph/9712318].

\bibitem{Lahde:1999ih}
T.~A.~Lahde, C.~J.~Nyfalt and D.~O.~Riska,
Nucl.\ Phys.\ A {\bf 674}, 141 (2000)
[arXiv:hep-ph/9908485].

\bibitem{DiPierro:2001uu}
M.~Di Pierro and E.~Eichten,
Phys.\ Rev.\ D {\bf 64}, 114004 (2001)
[arXiv:hep-ph/0104208].

\bibitem{Chao:2004nb}
K.~T.~Chao,
arXiv:hep-ph/0407091.

\bibitem{PDG2004}
S. Eidelman {\it et al.}, 
Phys. Lett. {\bf B}592, 1 (2004)

\bibitem{Barnes:2002mu}
T.~Barnes, N.~Black and P.~R.~Page,
Phys.\ Rev.\ D {\bf 68}, 054014 (2003).

\bibitem{Aston:1984ab}
D.~Aston {\it et al.},
Phys.\ Lett.\ B {\bf 149}, 258 (1984).

\bibitem{LeYaouanc:1977ux}
A.~Le Yaouanc, L.~Oliver, O.~Pene and J.~C.~Raynal,
Phys.\ Lett.\ B {\bf 71}, 397 (1977).

\bibitem{Micu:1968mk}
L.~Micu,
Nucl.\ Phys.\ B {\bf 10}, 521 (1969). 

\bibitem{LeYaouanc:1972ae}
A.~Le Yaouanc, L.~Oliver, O.~Pene and J.~C.~Raynal,
Phys.\ Rev.\ D {\bf 8}, 2223 (1973).

\bibitem{LeYaouanc:1973xz}
A.~Le Yaouanc, L.~Oliver, O.~Pene and J.~C.~Raynal,
Phys.\ Rev.\ D {\bf 9}, 1415 (1974).

\bibitem{LeYaouanc:1974ri}
A.~Le Yaouanc, L.~Oliver, O.~Pene and J.~C.~Raynal,
Phys.\ Rev.\ D {\bf 11}, 680 (1975).

\bibitem{LeYaouanc:1974mr}
A.~Le Yaouanc, L.~Oliver, O.~Pene and J.~C.~Raynal,
Phys.\ Rev.\ D {\bf 11}, 1272 (1975).

\bibitem{Ackleh:1996yt}
E.~S.~Ackleh, T.~Barnes and E.~S.~Swanson,
Phys.\ Rev.\ D {\bf 54} (1996) 6811.

\bibitem{Barnes:1996ff}
T.~Barnes, F.~E.~Close, P.~R.~Page and E.~S.~Swanson,
Phys.\ Rev.\ D {\bf 55}, 4157 (1997).

\bibitem{Capstick:2000qj}
S.~Capstick and W.~Roberts,
Prog.\ Part.\ Nucl.\ Phys.\  {\bf 45}, S241 (2000).

\bibitem{Capstick:kb}
S.~Capstick and W.~Roberts,
Phys.\ Rev.\ D {\bf 49}, 4570 (1994).

\bibitem{Godfrey:2003kg}
S.~Godfrey,
Phys.\ Lett.\ B {\bf 568}, 254 (2003)
[arXiv:hep-ph/0305122].

\bibitem{Godfrey:2004ya}
S.~Godfrey,
arXiv:hep-ph/0406228.

\bibitem{Barnes:1995hc}
T.~Barnes, F.~E.~Close and E.~S.~Swanson,
Phys.\ Rev.\ D {\bf 52}, 5242 (1995).

\bibitem{Bernard:1997ib}
C.~W.~Bernard {\it et al.}  [MILC Collaboration],
Phys.\ Rev.\ D {\bf 56}, 7039 (1997).

\bibitem{Lacock:1998be}
P.~Lacock and K.~Schilling  [TXL collaboration],
Nucl.\ Phys.\ Proc.\ Suppl.\  {\bf 73}, 261 (1999).

\bibitem{Luo:2002rz}
X.~Q.~Luo and Z.~H.~Mei,
Nucl.\ Phys.\ Proc.\ Suppl.\  {\bf 119}, 263 (2003).

\bibitem{Liao:2002rj}
X.~Liao and T.~Manke,
arXiv:hep-lat/0210030.

\bibitem{Adams:1998ff}
G.~S.~Adams {\it et al.}  [E852 Collaboration],
Phys.\ Rev.\ Lett.\  {\bf 81}, 5760 (1998).

\bibitem{Bernard:2003jd}
C.~Bernard {\it et al.},
Phys.\ Rev.\ D {\bf 68}, 074505 (2003).

\bibitem{Voloshin:1976ap}
M.~B.~Voloshin and L.~B.~Okun,
JETP Lett.\  {\bf 23}, 333 (1976)
[Pisma Zh.\ Eksp.\ Teor.\ Fiz.\  {\bf 23}, 369 (1976)].

\bibitem{DeRujula:1976qd}
A.~De Rujula, H.~Georgi and S.~L.~Glashow,
Phys.\ Rev.\ Lett.\  {\bf 38}, 317 (1977).

\bibitem{Tornqvist:1993ng}
N.~A.~Tornqvist,
Z.\ Phys.\ C {\bf 61}, 525 (1994);

\bibitem{Ericson:1993wy}
T.~E.~O.~Ericson and G.~Karl,
Phys.\ Lett.\ B {\bf 309}, 426 (1993).

\bibitem{Weinstein:1990gu}
J.~D.~Weinstein and N.~Isgur,
Phys.\ Rev.\ D {\bf 41}, 2236 (1990).

\bibitem{Dooley:1991bg}
K.~Dooley, E.~S.~Swanson and T.~Barnes,
Phys.\ Lett.\ B {\bf 275}, 478 (1992).

\bibitem{Szczepaniak:2003vy}
A.~P.~Szczepaniak,
Phys.\ Lett.\ B {\bf 567}, 23 (2003)
[arXiv:hep-ph/0305060].

\bibitem{Barnes:2003dj}
T.~Barnes, F.~E.~Close and H.~J.~Lipkin,
Phys.\ Rev.\ D {\bf 68}, 054006 (2003)
[arXiv:hep-ph/0305025].

\bibitem{vanBeveren:2004bz}
E.~van Beveren and G.~Rupp,
arXiv:hep-ph/0406242.

\bibitem{Tornqvist:2003na}
N.~A.~Tornqvist,
arXiv:hep-ph/0308277.

\bibitem{Close:2003sg}
F.~E.~Close and P.~R.~Page,
Phys.\ Lett.\ B {\bf 578}, 119 (2004)
[arXiv:hep-ph/0309253].

\bibitem{Braaten:2003he}
E.~Braaten and M.~Kusunoki,
Phys.\ Rev.\ D {\bf 69}, 074005 (2004)
[arXiv:hep-ph/0311147].

\bibitem{Swanson:2003tb}
E.~S.~Swanson,
Phys.\ Lett.\ B {\bf 588}, 189 (2004)
[arXiv:hep-ph/0311229].

\bibitem{Maiani:2004xg}
L.~Maiani, F.~Piccinini, A.~D.~Polosa and V.~Riquer,
arXiv:hep-ph/0407025.

\bibitem{Chen:2004dy}
Y.~Q.~Chen and X.~Q.~Li,
arXiv:hep-ph/0407062.

\bibitem{Nicolescu:2004in}
B.~Nicolescu and J.~P.~B.~de Melo,
arXiv:hep-ph/0407088.

\bibitem{LK}
H. Lipkin and M. Karliner, private communication.

\end{thebibliography}
\end{document}